\pdfoutput=1
\documentclass[11pt]{article}
\usepackage{emnlp2021}
\usepackage{times}
\usepackage{latexsym}
\usepackage[T1]{fontenc}
\usepackage[utf8]{inputenc}
\usepackage{microtype}
\usepackage{todonotes}
\usepackage{enumitem}
\usepackage{graphicx}
\usepackage{multirow}

\title{PoliWAM: An Exploration of a Large Scale Corpus of Political Discussions on WhatsApp Messenger}

\author{Vivek Srivastava \\
  TCS Research\\ Pune, Maharashtra, India \\
  \texttt{srivastava.vivek2@tcs.com} \\\And
  Mayank Singh \\
  IIT Gandhinagar\\ Gandhinagar, Gujarat, India \\
  \texttt{singh.mayank@iitgn.ac.in} \\}

\begin{document}
\maketitle
\begin{abstract}
WhatsApp Messenger is one of the most popular channels for spreading information with a current reach of more than 180 countries and 2 billion people. Its widespread usage has made it one of the most popular media for information propagation among the masses during any socially engaging event. In the recent past, several countries have witnessed its effectiveness and influence in political and social campaigns. We observe a high surge in information and propaganda flow during election campaigning. In this paper, we explore a high-quality large-scale user-generated dataset curated from WhatsApp comprising of 281 groups, 31,078 unique users, and 223,404 messages shared before, during, and after the Indian General Elections 2019, encompassing all major Indian political parties and leaders. In addition to the raw noisy user-generated data, we present a fine-grained annotated dataset of 3,848 messages that will be useful to understand the various dimensions of WhatsApp political campaigning. We present several complementary insights into the investigative and sensational news stories from the same period. Exploratory data analysis and experiments showcase several exciting results and future research opportunities. To facilitate reproducible research, we make the anonymized datasets available in the public domain. 
\end{abstract}

\section{Introduction}
In the last decade, the majority of the political parties around the world are heavily spending on social media engagement platforms like Facebook, Twitter, Quora, WhatsApp, and Sharechat for fast and secure information spread. WhatsApp Messenger (hereafter \textit{`WAM'}) is highly prevalent in 180 countries with installation over 90\% of devices~\cite{stat}. WAM allows users to send instant messages, photos, videos, and voice messages in addition to voice and video calls over a secure end-to-end encryption channel. However, data curation from WAM remains a challenging task owing to privacy concerns, stringent encryption strategies, and system requirements.

\noindent \textbf{WAM as Political Propaganda Tool:} Several investigative journalism stories~\cite{WhatsApp_investigative1,invest2,groups} suggest ever-increasing global penetration of WAM-based political propaganda and the resultant mass polarization effects. India, the second-most-populous country with 1.3 billion population, has witnessed similar trends~\cite{news1,news2} during the past two General Elections~\cite{whatsapp}. WAM has emerged as a primary leader for delivery for political messaging with 95\% of Android devices in India having WAM installation and maintaining a 75\% daily active users percentage~\cite{stat}. For the first time, complementing the existing investigative journalism stories, we present a scientific study to understand the WAM messaging patterns and spread in the political scenario. 

\noindent \textbf{WAM Data Curation:}
To adhere to the WAM's privacy policy~\cite{garimella2018whatapp}, we consider only WAM public groups where users willingly share messages with known and unknown people. The biggest challenge with restricting to public groups is a large number of fake and dubious groups and the unavailability of a standard in-app filter functionality to identify such groups. To the best of our information, none of the previous works on WAM data curation have discussed any methodologies for filtering irrelevant groups. One of the earliest works on WAM dataset curation~\cite{garimella2018whatapp} considers a list of all WAM public groups available online.  We claim that due to multilingualism coupled with code-mixing and multi-modal metadata, the genuine group identification problem is non-trivial. We present a novel manual group filtering strategy to filter fake and dubious groups by leveraging group metadata, i.e., group name, display picture, and the description. Owing to the user's privacy in the WAM groups, we release the anonymized dataset\footnote{\url{https://zenodo.org/record/4115660\#.YUSln54zZQI}}. In addition, we are also releasing a fine-grained annotated dataset of 3,848 messages (see Section \ref{sec: annotation}) for future research opportunities. The annotated dataset contains interesting fine-grained labels in four categories i.e., malicious activity, political orientation, political inclination, and message language.

\noindent \textbf{Our Contributions:} The main contributions are:
\begin{itemize}[noitemsep,nolistsep,leftmargin=*]
 \item We present a novel WAM group filtering strategy leveraging the group's metadata to filter the fake and dubious groups.
 \item We analyze a total of 281 public groups distributed among 26 political parties, including all the seven national political parties.
 \item In addition to the original anonymized dataset with 223,403 messages and 31,078 unique WAM users, we also release a publicly available fine-grained annotated dataset of 3,848 WAM messages with language, malicious activity, political orientation, and political inclination as labels.
 \item We present several interesting insights from the analysis of users and the message content. We establish several correlations with the claims and reports of news media and survey articles from the same period.
 \item We draw several interesting insights on co-occurrences of the political entities (leaders and parties) and social factors (agendas, religions, castes, and languages/ethnicity) from the linguistic perspective.
\end{itemize}

\section{Related Work}
People engage in various social media platforms such as Twitter, Facebook, etc., to discuss socially relevant topics such as politics. We witness a large volume of research focused on Twitter-based political discussions due to the easier availability of data and the large-scale involvement of the masses. For example, \cite{bovet2019influence} conducted a large-scale analysis of fake news propagation on Twitter during the 2016 U.S. presidential elections. \cite{tumasjan2010predicting} studied the usage of Twitter as a tool for political deliberation and election forecasting. They also used Twitter as a means to understand the political sentiment of the politicians and parties in the election campaigning. \cite{conover2011political} presented a study of the interaction between the network of users of different ideologies. They used Twitter as a medium to understand the political communication network. \cite{colleoni2014echo} studied the structure of political homophily between different political groups on Twitter to understand the public sphere and echo chamber effect. 

In contrast to Twitter-based studies, \cite{vitak2011s} studied student's involvement in the political discussions on Facebook for the 2008 U.S. presidential election.
Several studies looked at different offline modes of political discussions as well. For example, \cite{ansolabehere2006orientation} explored the effect of newspaper endorsement on the shift of vote margin. The largest circulation newspaper during the period of 1940--2002 is part of this experiment for the various U.S. elections. \cite{barrett2005picture} conducted a study to understand the visual perception of the voter by the candidate's photograph in the newspaper. \cite{wang2015forecasting} proposed an innovative methodology on non-representative polls to forecast elections in contrast to the survey methods. They experiment with the data from the Xbox gaming platform for the 2012 U.S. presidential election. \cite{ferreira2018analyzing} discuss the involvement of the ideological communities over 15 years. They used data on the public voting of Brazil and the U.S and study the polarization in the communities. \cite{caetano2018analyzing} presented an analysis and characterization of WAM messages at three different layers.

Recently, we observe a growing interest in understanding the information flow on WAM with political discussions in focus~\cite{resende2019information, resende2019analyzing, pang2020whatsapp, vermeer2021whatsapp, velasquez2021whatsapp, garimella2020images, saha2021short, javed2020first, yadav2020understanding}. In contrast to the existing works, we present a novel manual filtering strategy (see Section \ref{sec: dataset}) to identify the potential WAM groups. Our group identification strategy results in a high-quality dataset with less dubious groups. Also, to understand the various user and message characteristics, we present a fine-grained annotated dataset of 3,848 messages.

\section{Dataset}
\label{sec: dataset}
\begin{figure*}[!h]
\centering
\small{
\begin{tabular}{@{}c@{}c@{}c@{}}
  \includegraphics[width=0.45\linewidth]{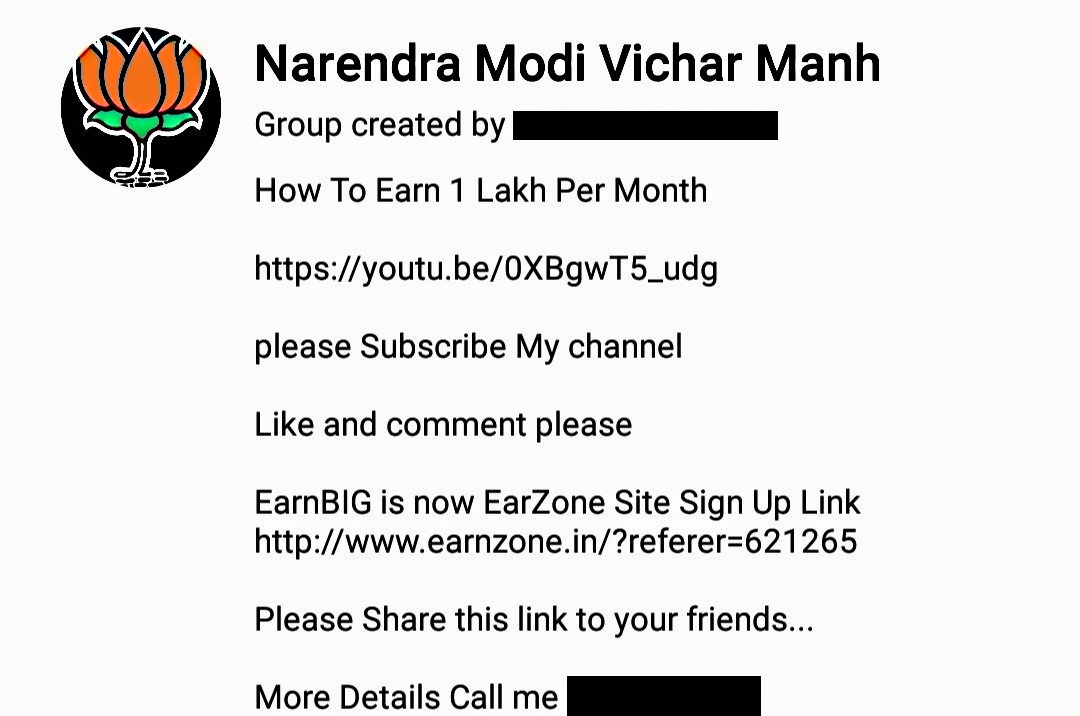} &
  \includegraphics[width=0.45\linewidth]{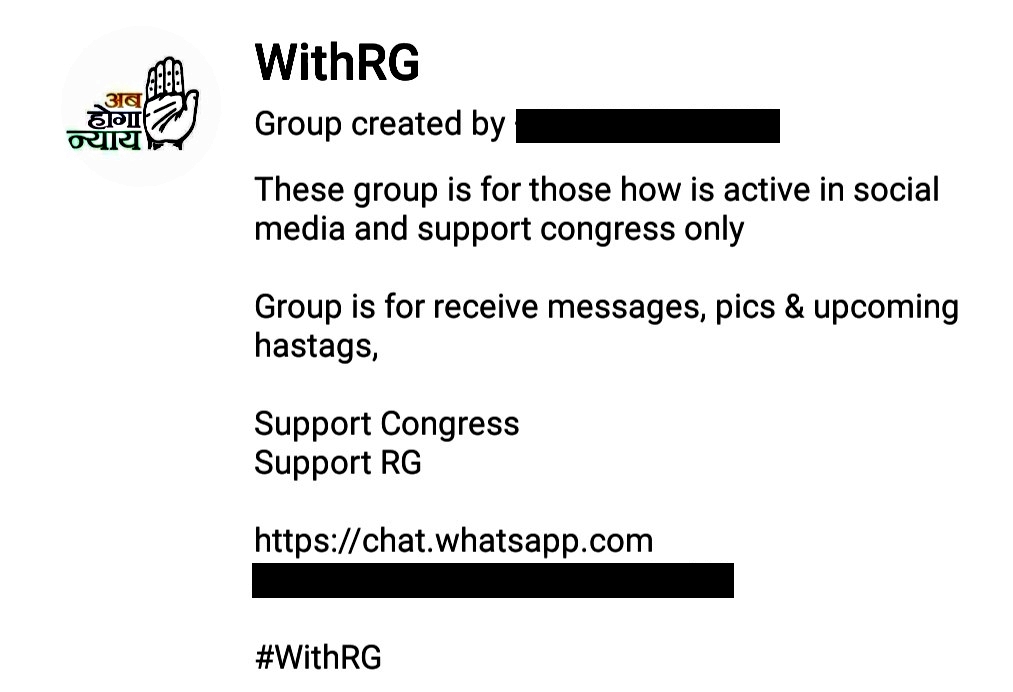} & \\
  (a) & (b) \\
\end{tabular}
}\caption{Example metadata snapshot of (a) a dubious and (b) a genuine WAM group. The personal information is anonymized.}
\label{fig:groups}
\end{figure*}

In contrast to other social media platforms, data curation from WAM public groups is a non-trivial task. Even, group identification itself is a challenging task as WAM does not support advanced content search feature.  Majority of the online forums/blogs\footnote{\url{https://www.opentechinfo.com/WhatsApp-groups/}}\footnote{\url{https://chatwhatsappgrouplink.blogspot.com/p/join-whatsapp-group-links.html?m=1}} comprises dubious public group links containing pornographic content, lucrative job, and lottery offers (see Figure \ref{fig:groups}). 

As the initial step of data curation, we identify and join the relevant groups. We construct a seed set of Indian political WAM groups (based on the names of political parties and their top leaders) from several forums/blogs and social media platforms like Google, Facebook, and Twitter. The seed set consists of 50 groups. The seed set is obtained after manual pre-processing of $\sim$600 groups. Further, we enrich the seed set by following groups that were shared within the followed groups. Overall, we obtained 2600 group links collected between January 19, 2019 -- May 19, 2019. Out of 2600, we only identify 281 ($\sim$11\%) relevant groups that are actively participating in Indian political discussions. Groups in which all the three metadata categories (i.e., name, display picture, and description) are indicative of the same political affiliation are considered genuine and relevant. 

\begin{figure*}[!t]
\centering
\small{
\begin{tabular}{@{}c@{}c@{}c@{}}
  \includegraphics[width=0.5\linewidth]{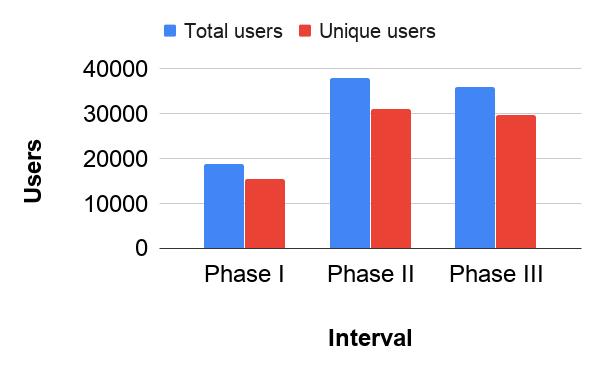} &
  \includegraphics[width=0.5\linewidth]{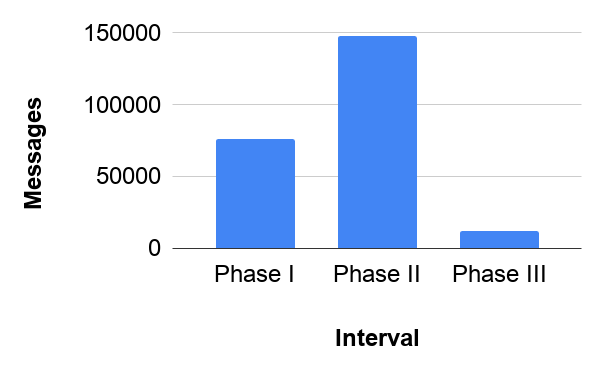} & \\
  (a) & (b) \\
\end{tabular}
}\caption{Status of different phases of data collection: (a) Number of total and unique users present and (b) Total messages shared in each phase.}
\label{comp}
\end{figure*}

For decryption of the WAM database stored locally on the mobile device, we use a cipher key and a database extractor tool\footnote{\label{tool} \url{https://forum.xda-developers.com/showthread.php?t=2770982}}. We divide the complete data collection into three phases:

\begin{itemize}[noitemsep,nolistsep,leftmargin=*]
    \item \textit{Before elections}: January 19, 2019 -- March 5, 2019
    \item \textit{Active campaigning phase}: March 6, 2019 -- May 19, 2019
    \item \textit{After elections}: May 20, 2019 -- June 15, 2019
\end{itemize}
Figure \ref{comp} presents user and content statistics at three different phases. We observe a sharp decline in the message count after elections. Similarly, before elections, we find a low user count. Thus, in the rest of the paper, we focus on the \textit{active campaigning phase}, where the maximum number of users has shared the highest number of messages.

\subsection{Metadata Analysis}

\begin{table*}[]
\resizebox{\hsize}{!}{
\begin{tabular}{|c|c||c|c||c|c||c|c||c|c|}
\hline
\textbf{Name} & \textbf{Count} & \textbf{Name} & \textbf{Count} & \textbf{Name} & \textbf{Count} & \textbf{Name} & \textbf{Count} & \textbf{Name} & \textbf{Count} \\ \hline
BJP           & 144            & AAP           & 20             & AIMIM         & 4              & RJD           & 3              & Shivsena      & 1              \\ \hline
INC           & 45             & BSP           & 9              & AIADMK        & 3              & AITC          & 3              & Others        & 19             \\ \hline
SP            & 21             & YSRCP         & 5              & NCP           & 3              & CPI(M)        & 1              &               &                \\ \hline
\end{tabular}}
\caption{Number of WAM groups affiliated to different political parties.}
\label{parties}
\end{table*}

Table \ref{parties} shows the representation of different political parties out of the total 281 groups. Each group is manually affiliated to a single political party based on the group name, display picture, and description. \textit{BJP} shows highest representation followed by \textit{INC} and \textit{SP}. All state-level parties, except \textit{AAP}, show poor representation. Several media reports~\cite{election_forecast1,election_forecast2} supports the insurgence of \textit{BJP} WAM groups to communicate with the various stakeholders. 

\begin{figure}[t]
\centering
\includegraphics[width=1\linewidth]{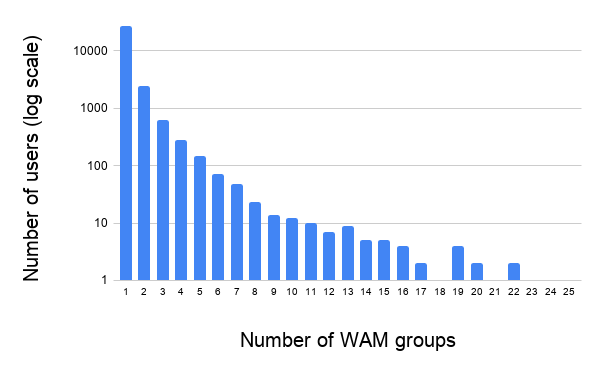}
\caption{User presence in multiple groups. One user is present in 25 groups. Seven users are part of 20 or more groups. }
\label{fig:user_presence}
\end{figure}

We found a total of 37,984 participants in 281 groups. Out of which, 31,078 (81.8\%) comprise unique users.  Figure~\ref{fig:user_presence} shows the user presence in multiple groups. Even though the majority of users (88.5\%) are members of a single group, we find instances where few users are members of more than 20 political groups. For instance, one user is a member of 25 groups, whereas seven users are part of 20 or more groups. In addition, we find that during \textit{active campaigning phase}, only a few (12.45\%) of these groups have group strength ${\le}$ 50 users. 63.34\% of groups have 100 or more members.


\subsubsection{Message formats}
In phase II, a total of 1,47,220 messages are shared, out of which 58,008 messages contain media files. Table~\ref{media} shows the distribution of media files. 67.59\% of the total media files are the image files in the JPEG format. This is consistent with the claims of several news articles that memes \cite{memes1,memes2} are a powerful tool to attack or praise individuals and parties during campaigning. Images (in JPEG format) and videos (in MP4 format) constitute 95.06\% of all the media files shared. Researchers conclude that image-based message sharing has extensively led to propaganda propagation~\cite{whatsapp_election}. For example, political parties/candidates use doctored images \cite{format,memes3} of articles from reputable news media sources to demean opponent political parties/candidates.  

\begin{table}[b]
\centering
\resizebox{\hsize}{!}{
\begin{tabular}{|c|c||c|c||c|c||c|c|} 
\hline
 \textbf{Type}                                                  & \textbf{Count}  & \textbf{Type}         & \textbf{Count} & \textbf{Type}                 & \textbf{Count}  & \textbf{Type}                           & \textbf{Count}   \\ 
\hline
 \textcolor[rgb]{0.502,0,0.502}{JPEG}  & 39,210           & \textcolor{red}{OGG}  & 371            & \textcolor{blue}{3GPP}        & 37              & \textcolor{red}{MP3}~                   & 1                \\ 
\hline
\textcolor{blue}{MP4}                                           & 15,934           & \textcolor{red}{MPEG} & 288            & \textcolor{red}{AAC}          & 39              & \textcolor[rgb]{0,0.4,0}{Octet-stream}  & 1                \\ 
\hline
\textcolor[rgb]{0.502,0,0.502}{WEBP}  & 1,221            & \textcolor{red}{MP4}  & 75            & \textcolor{red}{AMR}~         & 35              & \textcolor[rgb]{0,0.4,0}{Spreadsheets}  & 1                \\ 
\hline
\textcolor[rgb]{0,0.4,0}{PDF}                                   & 719             & APK                   & 66             & \textcolor[rgb]{0,0.4,0}{TXT} & 10              &    &                 \\
\hline
\end{tabular}}

\caption{(Best viewed in color) Distribution of shared media items. Blue, Red, Violet, Green and Black color represent video, audio, images, document and compressed files, respectively.}
\label{media}
\end{table}

\subsubsection{Group metadata}
\begin{figure}[t]
\centering
\begin{tabular}{@{}c@{}c@{}}
  \includegraphics[width=.45\hsize]{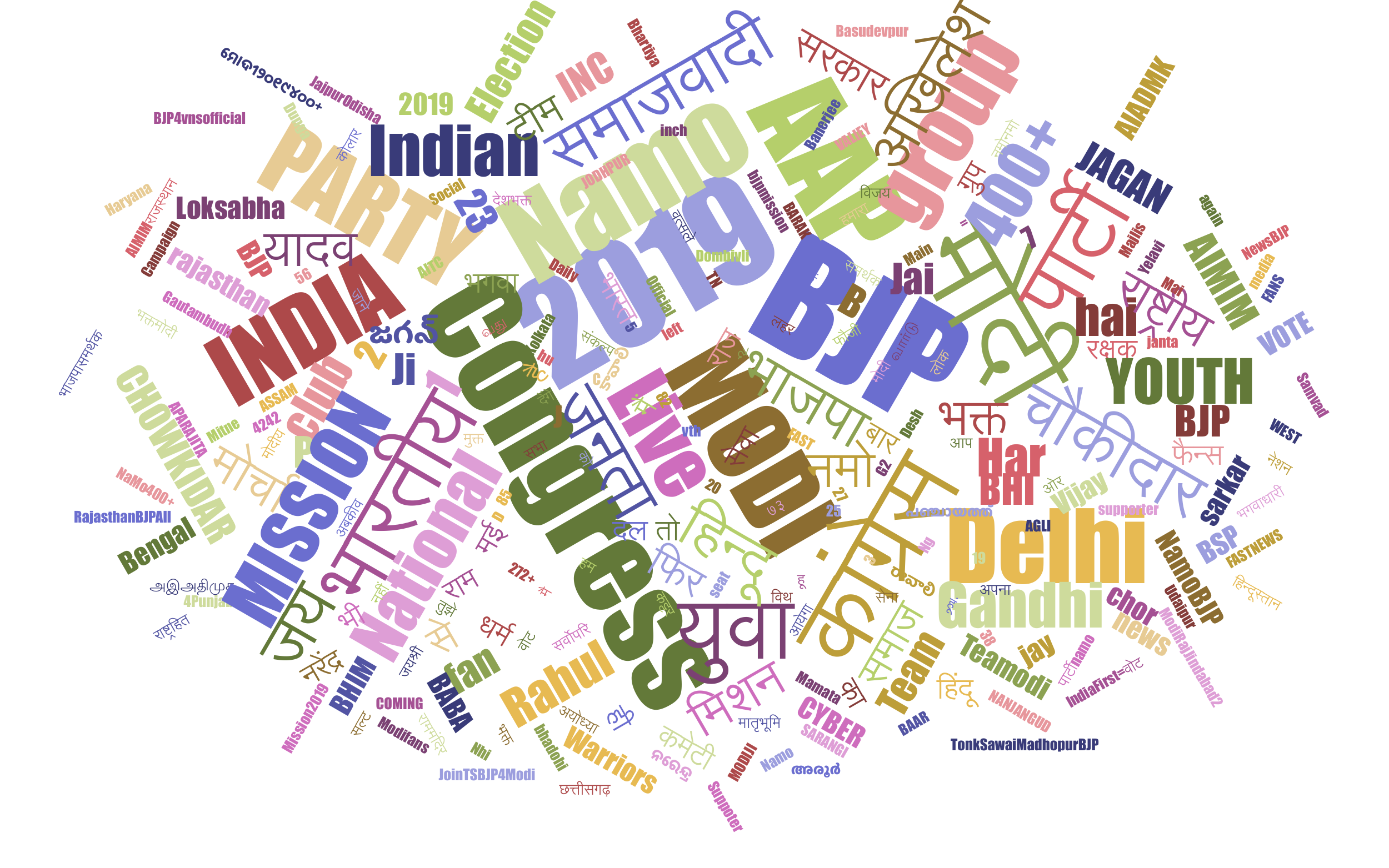} &
  \includegraphics[width=.45\hsize]{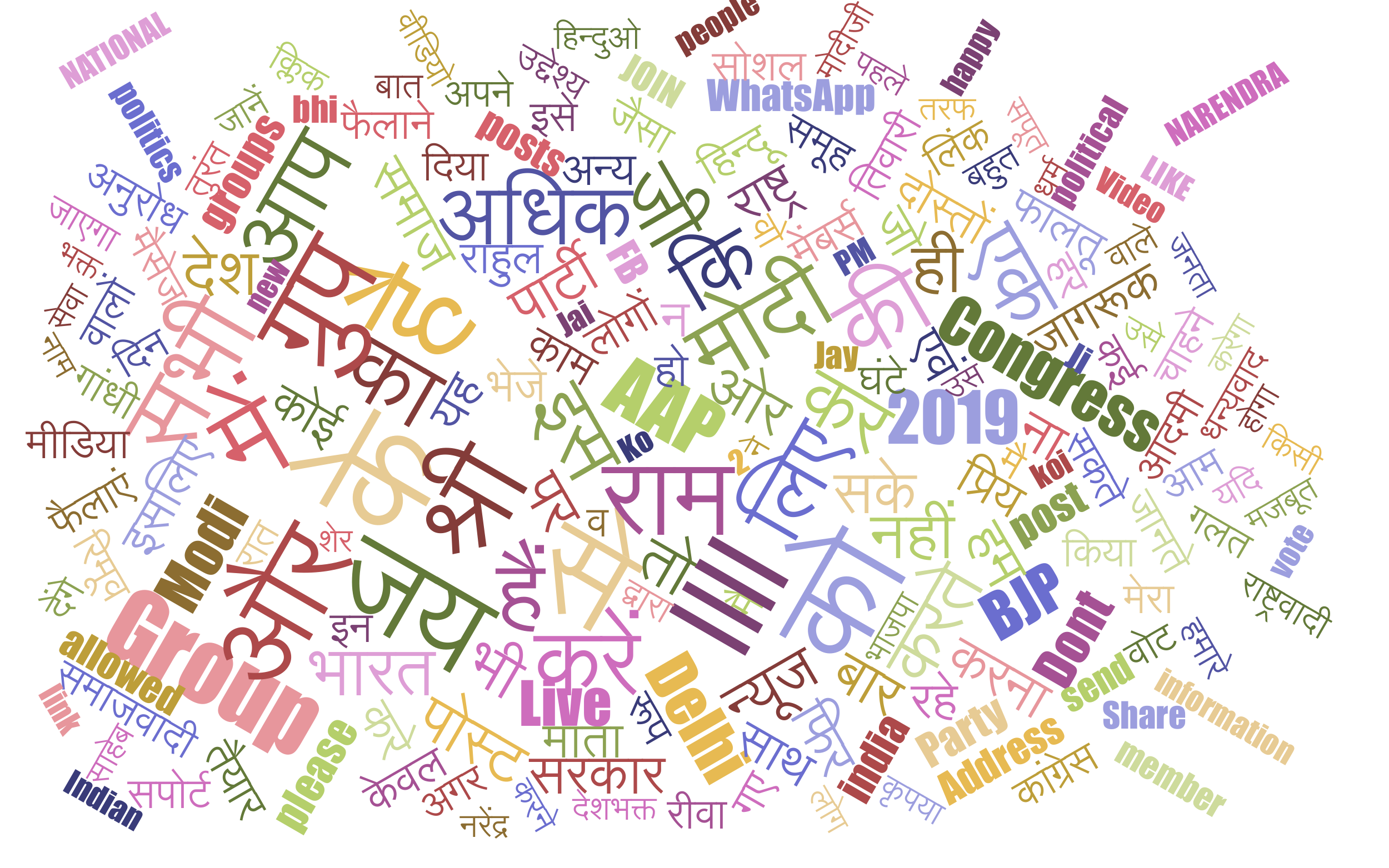} \\
  (a) & (b) \\
\end{tabular}
\caption{Word clouds of common phrases from group (a) names and (b) descriptions.} \label{fig:Word cloud1}
\end{figure}

Group metadata comprises a display picture, name, and description. Most of the group display pictures are the official party symbols, pictures of the top leaders/influencers, and religious symbols/idols. The textual content in group names and descriptions shows phrases used in social election campaigning. Figures ~\ref{fig:Word cloud1}(a) and~\ref{fig:Word cloud1}(b) show word clouds of the group names and group descriptions, respectively. We observe the most frequently used words to be in consistent with the claims \cite{words1,words2, words3} of various news and investigative studies. Several research and survey articles \cite{influencers1,influencers2} shows conflicting user opinion on the impact of slogans and influencers in the campaigning. Group names mainly consist of party names as English tokens. Group descriptions mostly contain Hindi language tokens. Thus, any group identification process, either manual or automatic, require a deep understanding of multiple languages.

\subsubsection{External link sharing}
Several website links are shared among group members. We find a total of 16,582 links comprising news, video, and social media websites. YouTube videos with 5,456 links are the most shared external links. This high number is in line with the claims of several studies \cite{youtube1,youtube2} indicating the high influence of YouTube videos in the campaigning. \textit{Facebook}, \textit{Twitter}, \textit{Dailyhunt} (Indian News mobile application), \textit{Jansatta} (Hindi daily for North India), \textit{NDTV} (an Indian television media company), etc., are the other popular websites with high number of links. The majority of the video links shared contain the news articles, speeches of the political leaders, and promotional content.  

\subsubsection{Contribution of top active users}
All users do not actively participate in the conversations.
The majority of the group members are passive consumers, while few users drive the topics of the conversation. In about 250 groups, the top 50\% active users share 80-100\% messages. Table~\ref{fig:top_usr_party} shows the contribution of top 1\%, 10\%, and 50\% active users in the groups of different political parties. AAP has the most contribution from the top 1\% users, whereas SP has the least contribution from top 1\%, 10\%, and 50\% users. BJP and INC, the top two political parties (based on the number of groups), show similar participation from the top 50\% active users. But, BJP leads INC in the participation of the top 1\% active users. This disproportionate distribution of user activity can be attributed to:
\begin{itemize}[noitemsep,nolistsep,leftmargin=*]
    \item Influencers: Presence of a few active participants/influencers in the groups who are possibly responsible for official campaigning of the party/candidates. 
    \item Neutral stance: Less participation from other members might indicate their neutral stance towards the ongoing campaigning, debates, and discussions \cite{invasion3}.
    \item Group invasion: Members from different political inclination join groups to share misinformation and fake news. Also, group invasion is possible to monitor (with least participation in discussions) the election campaigning of the opposition parties on WAM. Several claims \cite{invasion1,invasion2} have been made that indicates the high usage of WAM for such activities.
\end{itemize}

\begin{table}[!tbh]
\centering
\resizebox{\hsize}{!}{
\begin{tabular}{|c|c|c|c|c|c|c|}
\hline
                  & \textbf{BJP} & \textbf{INC} & \textbf{AAP} & \textbf{SP} & \textbf{BSP} & \textbf{Others} \\ \hline
\textbf{Top 1\%}  & 30.84        & 27.33        & 33.88        & 18.04       & 24.44        & 28.44           \\ \hline
\textbf{Top 10\%} & 67.20        & 64.63         & 69.82        & 52.17       & 56.20        & 65.23           \\ \hline
\textbf{Top 50\%} & 93.53        & 93.97        & 94.37        & 88.67       & 90.25        & 93.05           \\ \hline
\end{tabular}}
\caption{Percentage of messages shared by top 1\%, 10\%, and 50\% active users across different political parties.}
\label{fig:top_usr_party}
\end{table}

\subsubsection{Geographical analysis}
\begin{figure*}[t]
    \includegraphics[width=.30\textwidth]{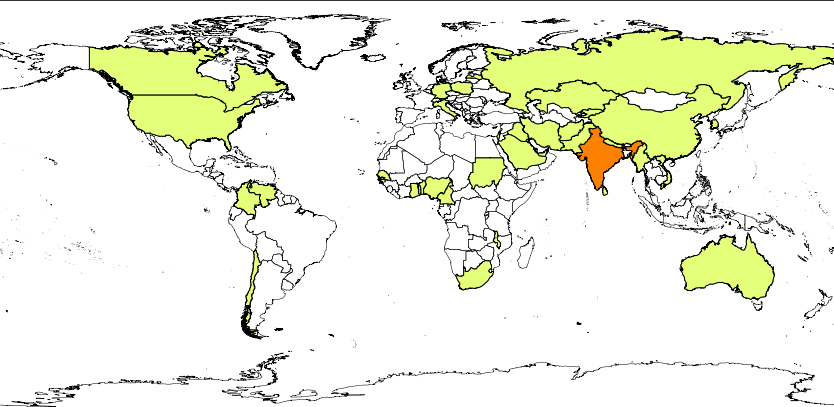}\hfill
    \includegraphics[width=.16\textwidth]{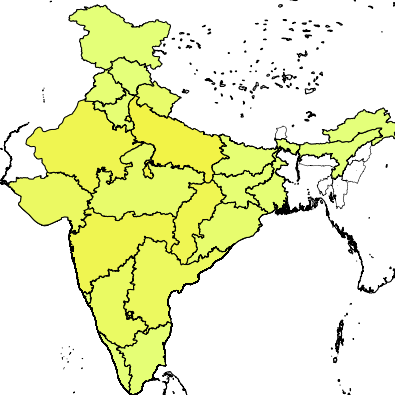}\hfill
    \includegraphics[width=.16\textwidth]{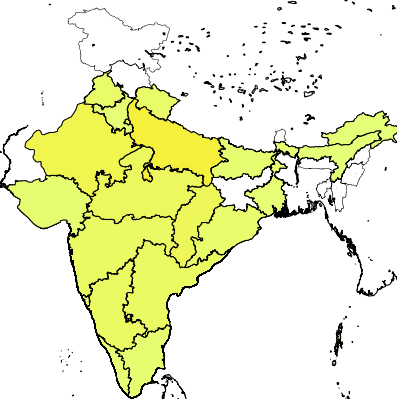}\hfill
    \includegraphics[height=60pt, width=.06\textwidth]{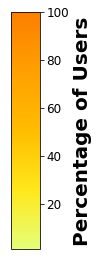}
    \caption{(Best viewed in color) (a) User locations in the World. (b) User locations in India. (c) Group administrator locations in India.}
    \label{map}
\end{figure*}

Next, we conduct the geographical analysis of group members based on their registered mobile number.\footnote{\url{https://www.searchyellowdirectory.com/reverse-phone/}} Surprisingly, we find users from 46 different countries discussing Indian politics. However, the majority of users belong to India. Figure ~\ref{map}(a) and ~\ref{map}(b) collectively show user locations outside and inside India. We also find that group administrators belong to five different countries (India, Pakistan, UAE, Latvia, and the USA). The above empirical findings confirm several claims about non-resident Indians participating in social media campaigns~\cite{NRI,nri1,nri2}. Figure ~\ref{map}(c) shows the locations of Indian group administrators. The majority of users and group administrators belong to northern and central India, strengthening the popular belief that WAM-based political campaigns are centrally managed by a team of IT experts headquartered in the Delhi-NCR region~\cite{ITCELL}.

\subsubsection{Group creation time}
Table~\ref{create} shows the group creation time of all the 281 groups. As we can see, a significant number (51.60\%) of groups are created in the first five months of the year 2019, which is consistent with claims made by several news articles~\cite{whatsapp_election,news1,news2,news3}.

\begin{table}[!tbh]
\centering
\resizebox{\hsize}{!}{
\begin{tabular}{|c|c||c|c||c|c|}
\hline
\textbf{Interval} & \textbf{Groups} & \textbf{Interval} & \textbf{Groups} & \textbf{Interval} & \textbf{Groups} \\ \hline
May'14-Dec'14      & 4               & May'16-Dec'16      & 11              & May'18-Dec'18     & 66              \\ \hline
Jan'15-Aug'15      & 2               & Jan'17-Aug'17      & 18              & Jan'19-May'19     & 145             \\ \hline
Sep'15-Apr'16      & 4               & Sep'17-Apr'18      & 31              &                   &                 \\ \hline
\end{tabular}}
\caption{Frequency of WAM group creation in different intervals.}
\label{create}
\end{table}

\subsubsection{Languages}
 Table~\ref{lang} presents language identification\footnote{\url{https://pypi.org/project/langdetect/}} statistics. A total of 31, 22, and 45 unique languages are used in writing group names, group descriptions, and messages, respectively. English and Hindi are the two most frequently used languages. The findings reiterate the challenges in conducting WAM-based user analysis, due to multilingualism, in highly diversified countries like India.  

\begin{table}[t]
\centering
\resizebox{\hsize}{!}{
\begin{tabular}{|l|c|}
\hline
&  \textbf{Top-5 languages in decreasing order of frequency}\\ \hline
\textbf{Names}&  \parbox[t]{70mm}{Hindi, English, Nepali, Marathi, Indonesian}\\ \hline
\textbf{Descriptions} & \parbox[t]{70mm}{Hindi, English, Nepali, Marathi, Tagalog}\\ \hline
\textbf{Messages}& \parbox[t]{70mm}{Hindi, English, Indonesian, Marathi, Somali} \\ \hline
\end{tabular}}
\caption{Top-5 languages found in group name and description and messages.}
\label{lang}
\end{table}

\begin{figure}[!tbh]
\centering
\includegraphics[width=1\linewidth]{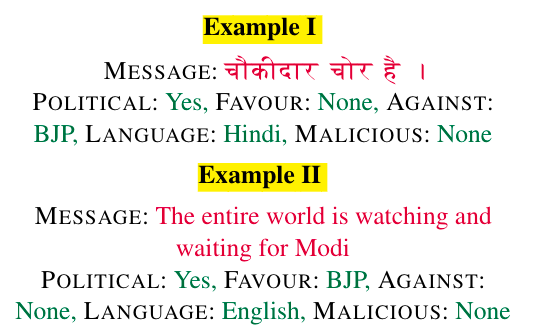}
\caption{Example annotation of the messages.}
\label{fig:example_problem}
\end{figure}

\subsection{Annotation}
\label{sec: annotation}
We construct a fine-grained manually annotated dataset of 3,848 messages. Eight annotators have performed annotation of the messages. All of the eight annotators are native Hindi speaker and proficient in speaking and writing English. Figure~\ref{fig:example_problem} shows the example annotation of the messages. Before annotation, we pre-process the original dataset to filter out irrelevant/noisy messages by removing non-textual data like hyperlinks, emoticons, images, and videos, duplicate messages, and messages with length less than five characters or more than 150 characters and keeping messages written in either of two scripts Roman or Devanagari. Pre-processing results in 48,474 messages. We randomly sample 3,848 messages from pre-processed data for annotation. Each annotator annotates 481 samples. Each message has a fine-grained annotation with the labels in the following categories:
\begin{table*}[h]
\resizebox{\hsize}{!}{
\begin{tabular}{|c|c|c|c|c||c|c||c|c|c|}
\hline
               & \multicolumn{4}{c|}{\textbf{Malicious activity}}                              & \multicolumn{2}{c|}{\textbf{Political orientation}} & \multicolumn{3}{c|}{\textbf{Language}}              \\ \hline
\textbf{}      & \textbf{Spam} & \textbf{Offensive} & \textbf{Advertisement} & \textbf{Others} & \textbf{Political}     & \textbf{Non-political}     & \textbf{Hindi} & \textbf{English} & \textbf{Others} \\ \hline
\textbf{Count} & 759           & 219                & 142                    & 971             & 1797                   & 2051                       & 2965          & 549             & 334             \\ \hline
\end{tabular}
}
\caption{Distribution of malicious activity, political orientation, and language in the annotated dataset.}
\label{orient}
\end{table*}

\begin{itemize}[noitemsep,nolistsep,leftmargin=*]
    \item \textbf{Malicious activity:} This category helps in identifying unusual and irrelevant activity within a group. A message is assigned one of the four labels in this category, i.e., spam, advertisement, offensive content, and others. Spam contains the textual messages that are irrelevant to the group and not part of any promotion of product/website/video/etc. Advertisement is specific to the promotional content. Others category contains the messages that are non-political and do not involve any other categories of malicious activities such as a non-political joke, historical content, personal conversation, etc. Others could be a set of messages with non-political and non-malicious content.
    Table \ref{orient} shows the malicious activity distribution. Spam is more prominent as compared to advertisement and offensive content. 
    \item \textbf{Political orientation:} Each message is assigned a binary label for the political orientation --- political or non-political. Table \ref{orient} shows the count of messages having political or non-political orientation. Even though the group identification process is completely manual with strict selection guidelines, we witness non-political messages surpassing political messages. We claim that without a strict selection criterion (as presented in earlier works~\cite{resende2019information,resende2019analyzing}), the insights will be highly noisy. 
    \item \textbf{Political inclination:} This category helps in identifying the political inclination (favor and against) of the messages based on their political affiliation. Similar to the categorization described in the previous section, we assign one of the seven labels (BJP, INC, SP, BSP, AAP, AITC, and Others) to each message. Table \ref{inclination} shows the inclination of the messages for different political parties. BJP is the most favored party based on the number of messages shared in favor. Whereas INC is the most targeted party.
    \item \textbf{Language:}  Each message is assigned one of the three labels Hindi, English, and Others. Table \ref{orient} shows the distribution of languages used in different messages. Hindi is the most preferred language.
\end{itemize}

\begin{table}[]
\centering
\resizebox{\hsize}{!}{
\begin{tabular}{|c|c|c|c|c|c|c|c|}
\hline
                 & \textbf{BJP} & \textbf{INC} & \textbf{AAP} & \textbf{SP} & \textbf{BSP} & \textbf{AITC} & \textbf{Others} \\ \hline
\textbf{Favour}  & 651          & 96           & 18           & 8           & 9            & 8             & 116             \\ \hline
\textbf{Against} & 298          & 600          & 44           & 42          & 44           & 52            & 114             \\ \hline
\end{tabular}}
\caption{Messages shared in favour and against of different political parties in the annotated dataset.}
\label{inclination}
\end{table}

\section{A Study on Political and Social Co-occurrences}
The majority of the political parties often declare the agendas for their election campaigns. These agendas are mostly driven by various factors that touch upon the needs, beliefs, and necessities of the stakeholders. During the election campaign, the information propagation on various social media platforms gets heavily influenced by these social factors.    

In this section, we present a discussion on the co-occurrence of two political entities (i.e., leaders and parties) and four social factors (i.e., agenda, religion, caste, and languages/ethnicity) in the dataset. Table \ref{tab: pol_soc} shows the fine-grained categorization of the political entities and the social factors. In our analysis, we select the six political parties (i.e. BJP, INC, AAP, SP, BSP, and AITC) and the most popular leader from each of these political parties (Narendra Modi--BJP, Rahul Gandhi--INC, Arvind Kejriwal--AAP, Akhilesh Yadav--SP, Mayawati--BSP, and Mamata Banarjee--AITC). The choice of social factors is motivated by the various influential topics being continuously reported to impact the political discourse in India.

To this end, we learn the word representation using FastText \cite{bojanowski2017enriching} on the textual data in the messages. We pre-filter the noisy content in the dataset i.e. emoticons, special characters, and hyperlinks. To learn the representation, we use Gensim\footnote{\url{https://radimrehurek.com/gensim/models/fasttext.html}} with Python interfaces and default parameters. English and Hindi are the two most popular languages in the dataset. The word representations from FastText has different representation for the same word in two languages. To address this challenge, we translate the political entities and the social factors in the English languages (Roman script) to the Hindi language (Devanagari script) using Google Translate\footnote{\url{https://translate.google.com/}}. To understand the co-occurrence of the political entities and social factors in the dataset, we compute the cosine similarity between them. For similarity computation, we use the words in both languages (i.e. English and Hindi). The similarity between political entity $w_1$ and social factor $w_2$ is given as:

\noindent
sim($w_1$, $w_2$) = max\{sim($hw_1$, $hw_2$), sim($hw_1$, $ew_2$), sim($ew_1$, $hw_2$), sim($ew_1$, $ew_2$)\},
where $hw_i$ and $ew_i$ represents the word representation of word $w_i$ in the Hindi and the English language respectively. 

For the entity/social factor with more than one word, we take the average of the word representation of the constituent words. Table \ref{tab:leader_sf} and \ref{tab: party_sf} presents the top-k (k=5 for `\textit{Agendas}', `\textit{Religions}', `\textit{Languages/Ethnicity}' and k=4 for `\textit{Castes}') most similar social factors for the political leaders and political parties respectively. We draw following major insights from the co-occurrence analysis:

\begin{table*}[!tbh]
\resizebox{\hsize}{!}{
\begin{tabular}{|c|c|c|}
\hline
\multirow{2}{*}{\textbf{Political}} &  Parties        &  BJP, INC, AAP, SP, BSP, AITC   \\ \cline{2-3} & Leaders        &  Narendra Modi, Rahul Gandhi, Arvind Kejriwal, Akhilesh Yadav, Mayawati, Mamata Banerjee  \\ \hline
\multirow{5}{*}{\textbf{Social}}    & Agendas        & \begin{tabular}[c]{@{}c@{}}Education, Development, Healthcare, Infrastructure, Manufacturing, Defense, Economy, \\  Transportation, Unemployment, Poverty, Terrorism, Literacy, Corruption, Inflation,\\ Taxation, Pollution, Digitization, Privatisation, Agriculture \end{tabular} \\ \cline{2-3} 
                           & 
Religion &  Hinduism, Islam, Sikhism, Christianity, Buddhism, Jainism \\ \cline{2-3}
& Castes & Forward Castes, Other Backward Class (OBC), Scheduled Castes (SC), Scheduled Tribes (ST) \\ \cline{2-3}
& \begin{tabular}[c]{@{}c@{}}Languages/\\ Ethnicity\end{tabular} & \begin{tabular}[c]{@{}c@{}}Hindi, Bengali, Marathi, Telugu, Tamil, Gujarati, Urdu, Kannada, Odia, Malayalam, Punjabi, \\Assamese, Maithili, Sanskrit \end{tabular}\\ 
                           
\hline
\end{tabular}}
\caption{Fine-grained categorization of the political entities and the social factors.}
\label{tab: pol_soc}
\end{table*}

\begin{table*}[!tbh]
\centering
\resizebox{\hsize}{!}{
\begin{tabular}{|l|c|c|c|c|}
\hline          & \textbf{Agendas}                                           & \textbf{Religions}                               & \textbf{Castes}                                 & \textbf{Languages/Ethnicity}                                                                \\ \hline
\textbf{Narendra Modi}   & \begin{tabular}[c]{@{}c@{}}Corruption, Literacy, Development, \\ Privatisation, Terrorism\end{tabular} & \begin{tabular}[c]{@{}c@{}}Sikhism, Buddhism, Jainism, \\ Hinduism, Islamism\end{tabular}      & \begin{tabular}[c]{@{}c@{}}FW, SC, \\ OBC, ST\end{tabular} & \begin{tabular}[c]{@{}c@{}}Odia, Assamese, Gujarati, \\ Sanskrit, Marathi\end{tabular}      \\ \hline
\textbf{Rahul Gandhi}    & \begin{tabular}[c]{@{}c@{}}Corruption, Unemployment, Literacy, \\ Development, Healthcare\end{tabular} & \begin{tabular}[c]{@{}c@{}}Sikhism, Buddhism, Islamism,\\  Jainism, Hinduism\end{tabular}      & \begin{tabular}[c]{@{}c@{}}FW, OBC, \\ SC, ST\end{tabular} & \begin{tabular}[c]{@{}c@{}}Odia, Assamese, Maithili, \\ Tamil, Marathi\end{tabular}         \\ \hline
\textbf{Arvind Kejriwal} & \begin{tabular}[c]{@{}c@{}}Corruption, Literacy, Taxation,\\  Terrorism, Healthcare\end{tabular}       & \begin{tabular}[c]{@{}c@{}}Sikhism, Buddhism, Islamism, \\ Jainism, Hinduism\end{tabular}      & \begin{tabular}[c]{@{}c@{}}OBC, FW, \\ SC, ST\end{tabular} & \begin{tabular}[c]{@{}c@{}}Odia, Assamese, Bengali, \\ Tamil, Maithili\end{tabular}         \\ \hline
\textbf{Akhilesh Yadav}  & \begin{tabular}[c]{@{}c@{}}Corruption, Development, Taxation, \\ Unemployment, Education\end{tabular}  & \begin{tabular}[c]{@{}c@{}}Sikhism, Hinduism, Jainism, \\ Islamism, Buddhism\end{tabular}      & \begin{tabular}[c]{@{}c@{}}FW, OBC, \\ ST, SC\end{tabular} & \begin{tabular}[c]{@{}c@{}}Maithili, Assamese, Odia, \\ Malayalam, Kannada\end{tabular}     \\ \hline
\textbf{Mayawati}        & \begin{tabular}[c]{@{}c@{}}Corruption, Literacy, Terrorism, \\ Taxation, Unemployment\end{tabular}     & \begin{tabular}[c]{@{}c@{}}Sikhism, Islamism, Hinduism, \\ Christianity, Buddhism\end{tabular} & \begin{tabular}[c]{@{}c@{}}FW, OBC, \\ ST, SC\end{tabular} & \begin{tabular}[c]{@{}c@{}}Odia, Assamese, Malayalam, \\ Maithili, Bengali\end{tabular}     \\ \hline
\textbf{Mamata Banerjee} & \begin{tabular}[c]{@{}c@{}}Literacy, Corruption, Poverty, \\ Development, Terrorism\end{tabular}       & \begin{tabular}[c]{@{}c@{}}Jainism, Sikhism, Buddhism, \\ Christianity, Hinduism\end{tabular}  & \begin{tabular}[c]{@{}c@{}}OBC, SC, \\ FW, ST\end{tabular} & \begin{tabular}[c]{@{}c@{}}Malayalam, Maithili, Marathi, \\ Sanskrit, Gujarati\end{tabular} \\ \hline
\end{tabular}}
\caption{Top-\textit{k} most similar social factors with each of the six political leaders. k=5 for the social factors `\textit{Agendas}', `\textit{Religions}', and `\textit{Languages/Ethnicity}'. k=4 for the social factor `\textit{Castes}'. Here, FW: Forward Castes , OBC: Other Backward Class, SC: Scheduled Castes, and ST: Scheduled Tribes.}
\label{tab:leader_sf}
\end{table*}

\begin{table*}[!tbh]
\centering
\resizebox{\hsize}{!}{
\begin{tabular}{|l|c|c|c|c|}
\hline
              & \textbf{Agendas}                                   & \textbf{Religions}                      & \textbf{Castes}                              & \textbf{Languages/Ethnicity}                                                           \\ \hline
\textbf{BJP}  & \begin{tabular}[c]{@{}c@{}}Corruption, Terrorism, Literacy, \\ Development, Unemployment\end{tabular}   & \begin{tabular}[c]{@{}c@{}}Hinduism, Islamism, Sikhism, \\ Jainism, Buddhism\end{tabular}     & \begin{tabular}[c]{@{}c@{}}OBC, FW, \\ SC, ST\end{tabular} & \begin{tabular}[c]{@{}c@{}}Assamese, Odia, Bengali, \\ Sanskrit, Hindi\end{tabular}    \\ \hline
\textbf{INC}  & \begin{tabular}[c]{@{}c@{}}Corruption, Terrorism, Unemployment, \\ Literacy, Development\end{tabular}   & \begin{tabular}[c]{@{}c@{}}Islamism, Hinduism, Sikhism, \\ Jainism, Buddhism\end{tabular}     & \begin{tabular}[c]{@{}c@{}}OBC, FW, \\ SC, ST\end{tabular} & \begin{tabular}[c]{@{}c@{}}Odia, Assamese, Bengali, \\ Hindi, Sanskrit\end{tabular}    \\ \hline
\textbf{AAP}  & \begin{tabular}[c]{@{}c@{}}Literacy, Taxation, Development, \\ Corruption, Unemployment\end{tabular}    & \begin{tabular}[c]{@{}c@{}}Hinduism, Sikhism, Islamism, \\ Christianity, Jainism\end{tabular} & \begin{tabular}[c]{@{}c@{}}OBC, FW, \\ SC, ST\end{tabular} & \begin{tabular}[c]{@{}c@{}}Hindi, Bengali, Sanskrit, \\ Maithili, Punjabi\end{tabular} \\ \hline
\textbf{SP}   & \begin{tabular}[c]{@{}c@{}}Unemployment, Corruption, Development, \\ Agriculture, Literacy\end{tabular} & \begin{tabular}[c]{@{}c@{}}Hinduism, Islamism, Christianity, \\ Jainism, Sikhism\end{tabular} & \begin{tabular}[c]{@{}c@{}}OBC, ST, \\ FW, SC\end{tabular} & \begin{tabular}[c]{@{}c@{}}Maithili, Odia, Assamese, \\ Gujarati, Punjabi\end{tabular} \\ \hline
\textbf{BSP}  & \begin{tabular}[c]{@{}c@{}}Unemployment, Corruption, Taxation, \\ Development, Education\end{tabular}   & \begin{tabular}[c]{@{}c@{}}Hinduism, Islamism, Jainism, \\ Sikhism, Christianity\end{tabular} & \begin{tabular}[c]{@{}c@{}}OBC, SC, \\ ST, FW\end{tabular} & \begin{tabular}[c]{@{}c@{}}Assamese, Odia, Bengali, \\ Gujarati, Maithili\end{tabular} \\ \hline
\textbf{AITC} & \begin{tabular}[c]{@{}c@{}}Digitization, Literacy, Corruption, \\ Development, Poverty\end{tabular}     & \begin{tabular}[c]{@{}c@{}}Christianity, Sikhism, Jainism, \\ Islamism, Hinduism\end{tabular} & \begin{tabular}[c]{@{}c@{}}ST, SC, \\ OBC, FW\end{tabular} & \begin{tabular}[c]{@{}c@{}}Odia, Assamese, Hindi, \\ Maithili, Punjabi\end{tabular}    \\ \hline
\end{tabular}}
\caption{Top-\textit{k} most similar social factors with each of the six political parties. k=5 for the social factors `\textit{Agendas}', `\textit{Religions}', and `\textit{Languages/Ethnicity}'. k=4 for the social factor `\textit{Castes}'. Here, FW: Forward Castes , OBC: Other Backward Class, SC: Scheduled Castes, and ST: Scheduled Tribes.}
\label{tab: party_sf}
\end{table*}

\begin{itemize}[noitemsep,nolistsep,leftmargin=*]
    \item `\textit{Corruption}', `\textit{Literacy}', and `\textit{Development}' are three most sought after agendas for political leaders and the parties including the ruling (i.e. BJP) and the opposition parties.
    \item `\textit{Privatisation}' is among the top-5 social factors for `\textit{Narendra Modi}'. No other leader or party has `\textit{Privatisation}' as the top-5 agendas.
    \item Though `\textit{Poverty}' is a major concern for India, only `\textit{Mamata Banarjee}' from `\textit{AITC}' has `\textit{Poverty}' as the top-5 agendas.
    \item Apart from `\textit{Corruption}', `\textit{Literacy}', and `\textit{Development}', the opposition leaders and parties shows high similarity with some fundamental issues such as `\textit{Healthcare}', `\textit{Unemployment}', `\textit{Education}', and `\textit{Taxation}'.
    \item Most of the political leaders shows low similarity with the two most debated religions in India (i.e. `\textit{Hinduism}' and `\textit{Islamism}'). The political parties shows the opposite behaviour with `\textit{Hinduism}' and `\textit{Islamism}' as the most associated religions.
    \item The order of similarity with `\textit{Hinduism}' and `\textit{Islamism}' for `\textit{BJP}' and `\textit{INC}' is opposite. `\textit{BJP}' is more similar with `\textit{Hinduism}' than `\textit{Islamism}' and vice-versa. 
    \item Most of the political leaders shows high similarity with the `\textit{Forward Castes}' whereas the political parties are more similar with the `\textit{Other Backward Class (OBC)}'.
    \item Apart from `\textit{Akhilesh Yadav}' and `\textit{Mayawati}', the `\textit{Scheduled Tribes (ST)}' remains the least similar with the political leaders. 
    \item The opposition leaders (i.e. leaders except `\textit{Narendra Modi}') shows high similarity with the Dravidian languages/ethnicity (i.e. `\textit{Tamil}', `\textit{Malayalam}', and `\textit{Kannada}').
    \item `\textit{Odia}' and `\textit{Assamese}' consistently shows high similarity with the political leaders and parties. The high similarity of `\textit{Odia}' can be attributed to the fact that state assembly elections of the state Odisha took place along with the Indian general elections 2019.
\end{itemize}

\section{Conclusion and Future Work}
In this paper, we present an exploration of the noisy user-generated data on WAM with a focus on Indian general elections 2019. In contrast to the existing works, we discuss a novel manual group filtering strategy to reduce the presence of noisy and dubious groups in the dataset. In addition, we present a fine-grained annotated dataset with multi-dimensional labels to understand various linguistic characteristics. Lastly, the co-occurrence analysis put forward several interesting insights from the Indian general elections 2019. In the future, the presented dataset would be useful in understanding various aspects (e.g., named-entities, POS tagging, language identification, etc.) of the noisy user generated textual data.

\bibliographystyle{acl_natbib}
\bibliography{custom}

\end{document}